\documentclass[sigconf]{acmart}

\usepackage[most]{tcolorbox}
\usepackage{subcaption} 
\usepackage{booktabs}
\tcbset{
  boxstyle/.style={
    colback=gray!5,          
    colframe=black!60,       
    boxrule=0.5pt,
    arc=2pt,
    left=5pt, right=5pt, top=5pt, bottom=5pt,
    colbacktitle=gray!25,      
    coltitle=black,         
    fonttitle=\bfseries,     
    borderline={0.5pt}{0pt}{black!60}, 
  },
}

\AtBeginDocument{%
  }

\setcopyright{acmlicensed}
\copyrightyear{2025}
\acmYear{2025}
\acmDOI{XXXXXXX.XXXXXXX}
\acmConference[Conference acronym 'XX]{Make sure to enter the correct
  conference title from your rights confirmation email}{June 03--05,
  2018}{Woodstock, NY}
\acmISBN{978-1-4503-XXXX-X/2018/06}

\begin{document}

\title{Bridging the Socio-Emotional Gap: The Functional Dimension of Human-AI Collaboration for Software Engineering}

\author{Lekshmi Murali Rani}
\affiliation{%
  \institution{Chalmers University of Technology and University of Gothenburg}
  \city{Gothenburg}
  \country{Sweden}
  \postcode{SE-41296}
}
\email{lekshmi@chalmers.se}

\author{Richard Berntsson Svensson}
\affiliation{%
  \institution{Chalmers University of Technology and University of Gothenburg}
  \city{Gothenburg}
  \country{Sweden}
  \postcode{SE-41296}
}
\email{richard@cse.gu.se}

\author{Robert Feldt}
\affiliation{%
  \institution{Chalmers University of Technology and University of Gothenburg}
  \city{Gothenburg}
  \country{Sweden}
  \postcode{SE-41296}
}
\email{robert.feldt@chalmers.se}

\renewcommand{\shortauthors}{Rani et al.}


\begin{abstract}

\textbf{Background.} As GenAI models are adopted to support the work of software engineers and their development teams,  understanding effective human-AI collaboration (HAIC) is increasingly important. Socio-emotional intelligence (SEI) enhances collaboration among human teammates, but its role in HAIC remains unclear. Current AI systems lack SEI capabilities that humans bring to teamwork, creating a potential gap in collaborative dynamics. \textbf{Objective.} In this study, we investigate how software practitioners perceive the socio-emotional gap in HAIC and what capabilities AI systems require for effective collaboration. \textbf{Method.} Through semi-structured interviews with 10 practitioners, we examine how they think about collaborating with human versus AI teammates, focusing on their SEI expectations and the AI capabilities they envision. \textbf{Results.} Results indicate that practitioners currently view AI models as intellectual teammates rather than social partners and expect fewer SEI attributes from them than from human teammates. However, they see the socio-emotional gap not as AI's failure to exhibit SEI traits, but as a functional gap in collaborative capabilities (AI's inability to negotiate responsibilities, adapt contextually, or maintain sustained partnerships). We introduce the concept of "functional equivalents": technical capabilities (internal cognition, contextual intelligence, adaptive learning, and collaborative intelligence) that achieve collaborative outcomes comparable to human SEI attributes. \textbf{Conclusion.} Our findings suggest that effective collaboration with AI for SE tasks may benefit from functional design rather than replicating human SEI traits for SE tasks, thereby redefining collaboration as functional alignment.

\end{abstract}




\maketitle

\section{Introduction}
The software engineering (SE) field is inherently collaborative, demanding effective communication, coordination, and teamwork among team members~\cite{lenberg2015behavioral,ahmadi2008survey}. Research shows that socio-emotional attributes such as emotional intelligence (EI), social skills, trust, team connection, clustering, and shared sense of identity play a crucial role in effective team collaboration that could lead to improved team performance and satisfaction in human teams~\cite{akgun2015antecedents,shameem2024impact,raki2016towards,datta2018does}. The integration of AI tools in SE tasks, specifically generative AI (GAI) tools, has introduced new dynamics and challenges, especially in human-AI collaboration (HAIC)~\cite{melegati2024exploring,hamza2024human,wang2020human}. When the effectiveness of SE tasks done through human-human (HH) collaboration can be enhanced through socio-emotional attributes~\cite{rezvani2019emotional,shameem2024impact,lenberg2015behavioral,madampe2022role}, there is a need to explore how these attributes function when humans collaborate with AI models that lack socio-emotional traits, especially in the SE contexts. 

AI systems lacking SEI attributes can hinder trust, empathy, and user engagement, ultimately reducing collaborative efficiency and productivity and limiting the potential benefits of AI integration~\cite{kolomaznik2024role,ulfert2024shaping}. Current research explores the relevance of SEI attributes such as trust, empathy, rapport, user engagement, and anthropomorphization in enhancing the effectiveness and productivity of HAIC ~\cite{chen2024feels,kolomaznik2024role}. However, to the best of our knowledge, the relevance of SEI attributes for effective HAIC in SE tasks has received limited attention, despite SE practitioners' unique awareness of the technical nuances and limitations inherent in AI tools, which shape their mental models on machine agency~\cite{sundar2020rise} and influence whether they treat AI as supporting tools or cognitive partners. While Kolomaznik et al. highlight the need for a framework shift in AI development to have a holistic approach in incorporating both technical proficiency and SEI~\cite{kolomaznik2024role}, it remains uncertain whether such SEI attributes are as relevant in SE as they are in other sectors such as healthcare, education, and customer service~\cite{melegati2024exploring}. 

This paper presents the findings of an empirical study that includes data collected through qualitative surveys (semi-structured interviews) with 10 practitioners (six in Phase 1 and four in Phase 2). It investigates how software practitioners currently perceive and adapt to the socio-emotional gap (characterized by AI's inability to understand context, recognize emotions, and maintain collaborative continuity) when working with AI teammates compared to human teammates. In particular, this study explores which aspects of SEI are most critical for effective collaboration in SE tasks and what challenges arise when AI systems lack these capabilities. The study then establishes how AI collaborative capabilities should evolve without synthetically replicating SEI traits, but by identifying functional equivalents that serve the same collaborative purposes as human SEI traits. The research aims to examine practitioners’ perceptions of SEI in human versus AI teammates and understand the socio-emotional gap in HAIC during SE tasks (RQ1) and identify how HAIC could evolve to enable effective collaboration (RQ2). The key contributions are (1) an empirical exploration of how technically proficient practitioners experience SEI differences between human and AI teammates in SE tasks and (2) a functional equivalents framework for technically proficient users in SE contexts that proposes technical mechanisms to support the same collaborative purposes as human SEI traits without replicating human emotions or social skills.  

The remainder of this paper is organized as follows. Section 2 presents background and related work. Section 3 describes the research methodology, detailing the research approach, data collection, data analysis, and limitations of the study. Section 4 presents the results and findings. Section 5 discusses the findings, with Section 6 concluding the paper.

\section{Background and Related Work}
Due to the interdisciplinary nature of this study, this section provides background on key concepts, including SEI traits and their relevance in SE tasks.

\noindent\textbf{Emotional and Social Intelligence Inventory}: The Emotional and Social Competency Inventory (ESCI) measures competencies that lead to superior performance~\cite{boyatzis2014emotional}. It includes emotional intelligence (EI) competencies (recognizing and using emotional information about oneself) and social intelligence (SI) competencies (recognizing and using emotional information about others)~\cite{seal2006fostering}, organized into four clusters: self-awareness and self-management (EI), and social awareness and relationship management (SI)~\cite{boyatzis2014emotional}. The ESCI-U expands this by including cognitive intelligence (CI), involving analytical thinking and information processing~\cite{boyatzis2014emotional}, making it particularly relevant for SE contexts requiring balanced technical problem-solving with interpersonal collaboration.

\noindent\textbf{SEI attributes in software team collaboration}: Research shows that SEI shapes software team dynamics through emotional awareness (enabling cooperation and collaborative learning~\cite{guevara2024socio,madampe2022role}) and emotional contagion( building trust and cohesiveness for effective communication~\cite{barczak2010antecedents}). 
Empirical studies show that teams with emotionally intelligent members exhibit better collaboration through effective emotion regulation and conflict management~\cite{cole2019building}, while those with higher extroversion, conscientiousness, openness, harmony, and autonomy show improved coordination and effectiveness~\cite{shameem2024impact}. These individual-level attributes interact with team-level social factors, including shared identity, connection, clustering behaviors, trust and empathy, to enhance performance, open communication, mutual understanding and project success~\cite{datta2018does,akgun2015antecedents,cheng2016trust}. These studies highlight the importance of SEI in human SE teams but do not address their relevance in HAI teams. 

\noindent\textbf{SEI attributes enhancing SE decision-making}: Research shows that SEI attributes such as EI, trust, motivation, and social awareness enhance decision-making in SE contexts~\cite{rodrigues2024relationship,kay2018self,calefato2011augmenting}, with social awareness particularly supporting knowledge sharing while fostering trust, communication and coordination~\cite{calefato2011augmenting}. High EI enables understanding emotional consequences of decisions, leading to sustainable, informed, and balanced outcomes, which is critical as SE decisions depend on both technical information and interpersonal dynamics~\cite{rodrigues2024relationship}. Additionally, developers' emotional responses to requirement changes impact their cognition, productivity, and decision-making, emphasizing the need to understand emotions to support informed decisions~\cite{madampe2022emotional}. As SEI enhances SE decision-making in human teams, it is essential to study its relevance in HAI teams.

\noindent\textbf{Socio-emotional aspects in HAIC}: Existing studies show collaborative efficiency improves when AI systems adapt to humans' emotional and cognitive needs~\cite{kolomaznik2024role}. Key SEI attributes in HAIC include trust, empathy, rapport, user engagement, and anthropomorphization~\cite{kolomaznik2024role}. AI systems must understand and respond to human emotions to build trust and create reliability~\cite{mallick2024you}. Anthropomorphization, which is the attribution of human-like characteristics to non-human entities (e.g., AI models), improves acceptance and collaboration~\cite{onnasch2021impact,epley2007seeing}. SEI attributes in HAIC improve shared mental models for better collaborative performance~\cite{meyer2022process}, increase engagement for more productive interactions~\cite{mallick2024you,kolomaznik2024role}, and reduce decision-making conflict due to the ability of the AI system to comprehend SEI aspects~\cite{ferrada2024emotions}. Though these studies highlight the relevance of SEI in HAIC, they do not account for domain-specific differences or user expertise.

While existing research establishes SEI importance in human software teams and general HAI, a critical gap remains in the specific SEI aspects that enhance HAIC in the SE domain. Given that SEI attributes enhance HH collaboration effectiveness, there is a need to explore these attributes when humans collaborate with AI models lacking SEI traits in SE contexts. Software practitioners' technical expertise and awareness of AI limitations may fundamentally alter their perceptions of the need for SEI traits in AI tools. This study thus investigates the relevance of SEI traits in HAIC, where humans possess high technical competence to understand AI tool limitations.

\section{Research Methodology}
This section provides an overview of the methodological approach of this study, including research questions, study design, data collection, and analysis.

\noindent\textbf{Research approach:} This study uses an exploratory knowledge-seeking research approach~\cite{stol2018abc,wohlin2015towards} to understand the socio-emotional gap in HAIC and explore resulting challenges in SE decision-making, from the perspective of technically proficient users (i.e., software practitioners) who collaborate with AI models/tools. The research uses inductive reasoning to infer theoretical concepts and patterns from the data collected to develop theoretical conclusions~\cite{wohlin2015towards}. Through exploratory and descriptive approaches~\cite{wohlin2015towards}, the study examined the problem area of the socio-emotional gap, identifying practitioners' perspectives about it, describing specific SEI aspects that influence collaboration in HAIC, along with the associated challenges and mitigation strategies in SE tasks. Qualitative methods (interviews) were used in this study to uncover deeper processes in individuals and HAI teams, gaining an understanding of what individuals experience and how they interpret their experiences and understanding of how these processes unfold over time~\cite{bluhm2011qualitative}. Two research questions guided this research:

\begin{tcolorbox}[boxstyle,boxsep=2pt,left=3pt,right=3pt,top=3pt,bottom=3pt,before skip=6pt,after skip=6pt,title={Research Questions}]

\textbf{RQ1:} What is the perception of software practitioners on the socio-emotional gap in HAIC for SE?\\
\textbf{RQ2:} How should HAIC capabilities evolve to enable effective collaboration in the context of the socio-emotional gap?

\end{tcolorbox}

\noindent\textbf{Study design and context:} Given the aim of this study, in Phase 1, we conducted six interviews to answer the research questions. The sample size of six participants was based on the principle of information power~\cite{malterud2016sample}, which indicates that studies with narrow aims, specific participants, an established theoretical framework, and in-depth discussion require fewer participants. The focused research questions and the depth of semi-structured interviews based on the ESCI scale provided rich, detailed data sufficient for thematic analysis. The first three interviews (P1-P3) generated all the seven major themes. Interviews 4 and 5 provided additional supporting evidence to these themes while Interview 6 confirmed saturation, contributing only to existing themes thus stabilizing the thematic structure. Based on the results from Phase 1, a functional equivalents framework was developed by mapping ESCI traits to collaborative capabilities in AI models. To validate this framework, four additional interviews were conducted in Phase 2. The results and findings of Phases 1 and 2 are presented in \autoref{sec:results}. Thematic saturation was monitored throughout data collection to ensure that no new themes emerged from the final interviews. 

\noindent\textbf{Participant selection and characteristics:} Participants were recruited from LinkedIn, industry contacts, and professional networks based on the following inclusion criteria: professional role in SE, minimum three year of SE experience, minimum 1 year active use and experience in Generative AI tools for SE tasks. Purposive sampling ensured diversity in roles, experience levels, and types of AI tools used to capture the range of perspectives in this qualitative research~\cite{bluhm2011qualitative}. Participants reported using a diverse range of AI tools, including coding assistants (GitHub Copilot, Cursor) and conversational AI models (ChatGPT, Claude, Gemini), for code generation, debugging, technical documentation, and brainstorming requirements.\autoref{participantdetails} presents the participant details. Phase 1 included six software practitioners (P1-P6) with technical experience ranging from 3 to 11 years and AI experience ranging from 1 to 3 years. Phase 2 included four additional practitioners (VP1-VP4) with technical experience ranging from 4 to 10 years and AI experience ranging from 2 to 3 years.

\begin{table}[ht]
  \caption{Participant Details}
  \label{participantdetails}
  \centering

  \begin{subtable}[t]{0.48\textwidth}
    \centering
    \caption{Phase 1 Interview Participants}
    \label{tab:mainparticipants}
    \begin{tabular}{cccc}
      \toprule
      ID & Job Role & Tech. exp. & AI exp. \\
      \midrule
      P1 & Test automation engineer & 8 years & 1 year \\
      P2 & Software Engineer         & 11 years & 1 year \\
      P3 & Python Developer          & 9 years & 3 years \\
      P4 & Senior Engineer           & 5 years & 2 years \\
      P5 & Researcher                & 3 years & 2 years \\
      P6 & Senior Technical Consultant & 5 years & 1.5 years \\
      \bottomrule
    \end{tabular}
  \end{subtable}
  \hfill
  \begin{subtable}[t]{0.48\textwidth}
    \centering
    \caption{Phase 2 Interview Participants}
    \label{tab:validationparticipants}
    \begin{tabular}{cccc}
      \toprule
      ID & Job Role & Tech. exp. & AI exp. \\
      \midrule
      VP1 & Software Engineer         & 10 years & 2 years \\
      VP2 & Machine Learning Engineer & 6 years  & 3 years \\
      VP3 & Industrial Researcher     & 4 years  & 3 years \\
      VP4 & Software Engineer         & 5 years  & 3 years \\
      \bottomrule
    \end{tabular}
  \end{subtable}

\end{table}

\noindent\textbf{Data Collection:} Based on a comprehensive literature review of SEI concepts~\cite{boyatzis2014emotional}, a semi-structured interview instrument was developed for the Phase 1 interview. Phase 1 interview explored participants' perceptions of AI effectiveness and decision-making support, their expectations of SEI traits in human versus AI teammates using the ESCI framework, and the challenges caused by the socio-emotional gap in HAIC, along with their envisioned improvements in AI collaborative capabilities. The interview instrument was designed with the idea that there must be flexibility in the process, while it must follow a certain level of structure. The SI, EI, and CI traits in the interview instrument were adopted from the ESCI-U instrument~\cite{boyatzis2014emotional} due to their relevance in collaborative team setups. The ESCI-U framework was used as an input for the structured vocabulary of socio-emotional competencies that could help participants compare their collaborative expectations from human and AI teammates~\cite{boyatzis2014emotional}. The Phase 2 interview instrument was developed using inputs from the functional equivalents framework and vocabulary of socio-emotional competencies. Phase 2 validated the relevance of the four functional equivalents (derived from Phase 1 findings) through specific SE use cases and scenarios. Both interview instruments can be found in Zenodo~\cite{anonymous_2025}. A pilot interview was conducted with one software practitioner to test the interview instruments. The interview instruments were revised based on feedback, and the pilot data were excluded from the final analysis. 

\begin{figure*}[t]
    \centering
    \includegraphics[width=0.80\textwidth,height=0.25\textheight, keepaspectratio]{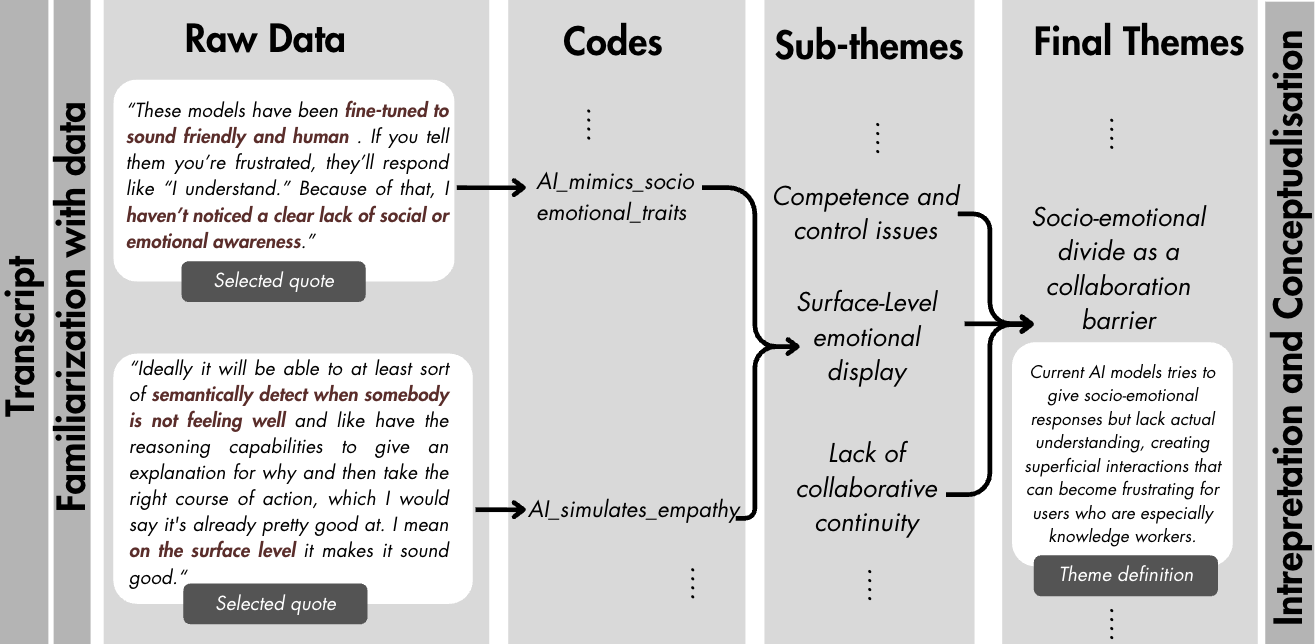}
    \Description{A visual summary of the thematic analysis process, showing how initial codes were refined into final themes following Braun and Clarke’s six-phase approach.}
    \caption{Thematic analysis process showing final theme development using Braun and Clarke’s six-phase approach~\cite{braun2006using}.}
    \label{thematic_analysis}
\end{figure*}

Each interview lasted 40-50 minutes and was conducted online using Microsoft Teams, with recordings transcribed in parallel for analysis. Informed consent for recording was obtained from participants, who were informed of their right to withdraw without consequences. To ensure confidentiality about participant information, participant IDs were used, and all identifying information was removed from the data during data cleaning. Participants were informed about data usage in research publications. The first author, who has eight years of IT/education experience and interdisciplinary training in psychology and SE, conducted interviews. This background balanced potential bias toward technical framing while enabling familiarity with SE practices and exploratory openness to practitioners' AI experiences\cite{lenberg2024qualitative}.

\noindent\textbf{Data Analysis:} Thematic analysis was used to systematically identify patterns and themes from the interview data (Phases 1 and 2) using Braun and Clarke's six-phase approach~\cite{braun2006using}, including the qualitative coding practices~\cite{bluhm2011qualitative}. After familiarizing with the data by reading the transcript multiple times, open coding was used to identify the initial categories from the interview data, followed by axial coding to find the relationship between various identified themes~\cite{bluhm2011qualitative}. Through constant comparison across interviews, sub-themes and themes were reviewed iteratively to ensure coherence and distinctiveness. After this refinement process, initial codes, sub-themes, and final themes were consolidated with clear definitions and representative participant quotes. Although coded data distribution was uneven across participants (e.g., P5 and P1), all themes were supported by atleast three participants indicating shared patterns rather than individual views. The \autoref{thematic_analysis} shows an example of the thematic coding process in Phase 1 data analysis. Thematic analysis of Phase 1 interview answered RQ1 and RQ2, and informed the development of functional equivalents that translate ESCI framework competencies into AI collaborative capabilities (\autoref{subsec:functionalequivalents}). This framework was validated through four additional interviews, with validation results presented in \autoref{subsec:validation}. Phase 2 thematic analysis was also conducted exactly in the same way as Phase 1. Findings from Phase 2 were used to refine theme descriptions and identify missing perspectives to improve the overall robustness of the findings.

\noindent\textbf{Limitations:} \textit{Credibility.} The interdisciplinary nature of HAIC for technical tasks introduces credibility challenges, as practitioners may interpret SEI traits differently in AI collaboration contexts, and interview questions may not fully capture the complexity of the socio-emotional gaps in HAIC. Additionally, the subjectivity in qualitative analysis introduces challenges in establishing credible interpretations. This was mitigated by providing clear definitions of SI, EI, and CI, with clear instructions on the context of the study, framing the study within an established framework (ESCI-U), and refining interview protocols through a pilot study. In addition, we critically examined the assumptions used in the research and discussed alternate interpretations. Practitioners' perceptions reflect current AI capabilities and their exposure to existing tools, but as AI technology evolves, these perceptions may change. To enhance credibility, two-phase data collection was done where Phase 2 validates Phase 1 findings, followed by structured thematic analysis to ensure concrete examples instead of abstract claims.\textit{Transferability.} The small sample size (n=10), while appropriate for exploratory qualitative research~\cite{bluhm2011qualitative}, limits the breadth of perspectives despite monitoring thematic saturation. Most participants' limited AI experience (1-3 years) may be insufficient for mature collaborative judgment. Practitioners may also show domain-specific skepticism and stated preferences may reflect adaptation to current AI limitations rather than fundamental needs.  We addressed these constraints through purposive sampling that ensured diversity across participant roles, experience levels, and AI tools. They may also exhibit unique skepticism compared to other technical populations, potentially caused by perceived professional threats.  While findings are context-specific, they may be transferable to similar settings where technically skilled practitioners collaborate with AI tools for knowledge work, but they may be limited in contexts with differing AI maturity levels and non-technical users. As AI gets more advanced capabilities, the functional-emotional distinction may blur. Practitioners currently prioritizing functional attributes might accept emotionally responsive interfaces if technical performance improves. Current preferences may reflect adaptation to AI's limited emotional intelligence rather than inherent human needs. \textit{Resonance.} Findings were not formally checked with Phase 1 participants, which may limit direct confirmation that the findings fully reflected their experiences. However, the validation interviews with additional practitioners supported the framework's relevance.

\section{Results and Findings}
\label{sec:results}

\begin{figure*}[t]
    \centering
    \includegraphics[width=0.7\textwidth, keepaspectratio]{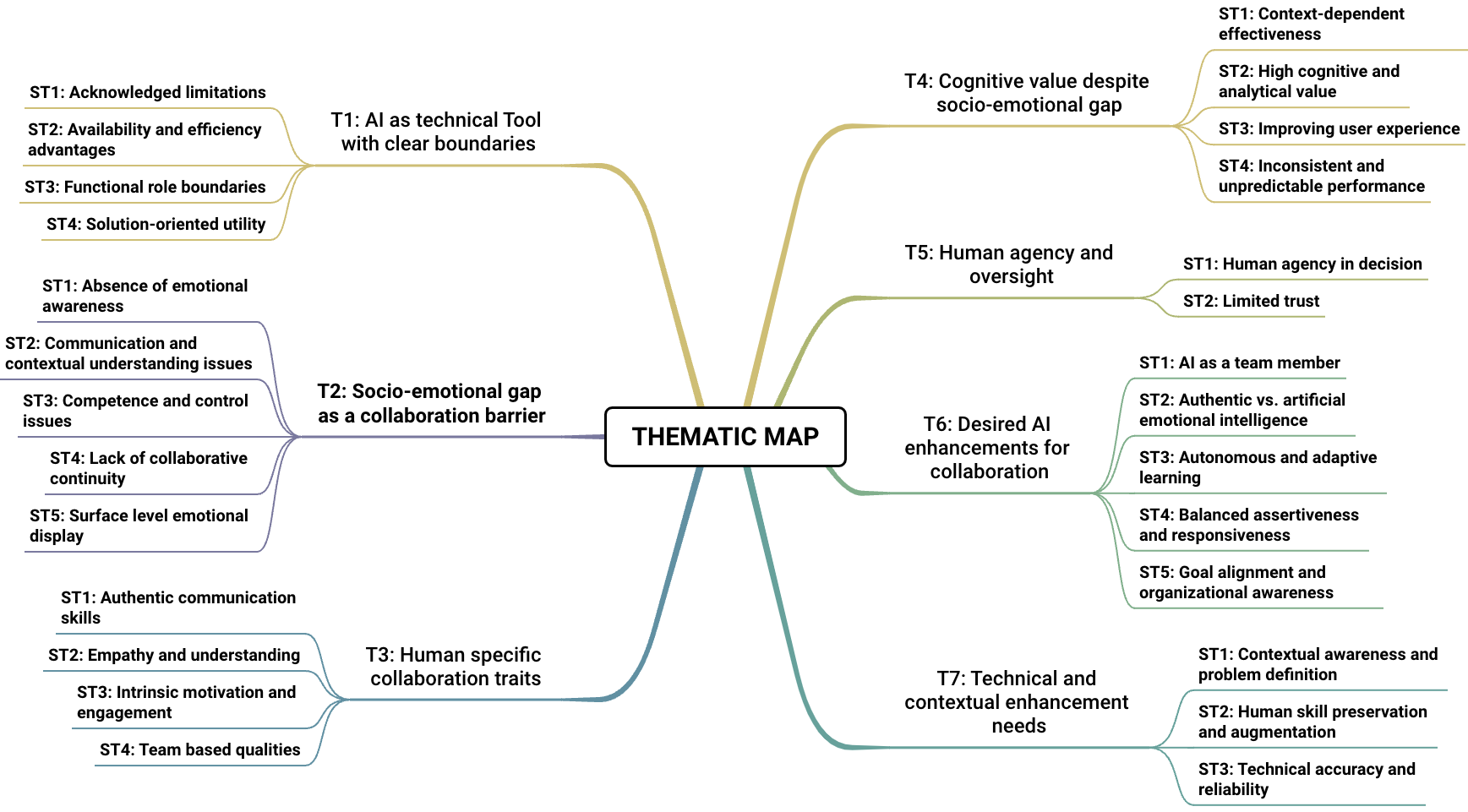}
    \Description{A thematic map showing seven themes. Themes 1–5 represent practitioners’ perceptions of the socio-emotional gap in human–AI collaboration, and Themes 6–7 depict the desired evolution of AI capabilities.}
    \caption{Thematic map showing the seven themes addressing practitioners' perceptions of the socio-emotional gap in HAIC (RQ1: Themes 1–5) and desired evolution of AI capabilities (RQ2: Themes 6–7).}
    \label{thematic_map}
\end{figure*}

This section presents the results of RQ1 (\autoref{subsec:RQ1}) and RQ2 (\autoref{subsec:RQ2}), followed by the functional equivalents framework (\autoref{subsec:functionalequivalents}) and framework validation (\autoref{subsec:validation}). Thematic analysis generated seven themes that capture practitioners' perceptions of the socio-emotional gap in HAIC and their visions for how AI capabilities should evolve to address the current collaboration challenges, as shown in the thematic map (\autoref{thematic_map}). The findings are organized around two research questions and supported by participant quotes. Themes 1–5 address RQ1, focusing on practitioners' perceptions of AI's current roles, socio-emotional limitations, and human collaborative capacities. Themes 6–7 address RQ2, exploring practitioners' visions for the evolution of collaborative, technical, and contextual dimensions in AI tools they use. Themes and sub-themes are labeled as T and ST, respectively, such that T–ST refers to Subtheme ST under Theme T (e.g., T1–ST3 denotes the third subtheme of Theme 1).

\subsection{RQ1: Practitioners' perceptions of the socio-emotional gap}
\label{subsec:RQ1}
Thematic analysis showed how practitioners currently position AI as a technical tool with clear boundaries (T1) and identify specific socio-emotional deficits that affect collaboration (T2). They compare AI limitations with distinctively human collaborative capacities (T3), while acknowledging AI's cognitive value despite these gaps (T4). This recognition of AI's limitations reinforces the need for human oversight and control (T5). 

\noindent\textbf{T1.AI as technical tool with clear boundaries:}
Practitioners view AI as a technical assistant with well-defined limitations, maintaining strict boundaries between AI capabilities and human collaboration needs. This perception creates clear functional boundaries between HH collaboration and HAIC. Practitioners recognize that AI possesses specific technical capabilities with limited scope for collaborative flexibility (T1-ST3). P2 highlight this: "\textit{If I have a deadline..., I can convey my team...to split the work and to complete that by tomorrow.}" This indicates that practitioners maintain realistic expectations about AI's role and understand that it cannot engage in true work distribution or collaborative problem-solving, nor do they expect it to. AI tools are primarily valued for providing answers and solutions to specific technical problems (T1-ST4). P4 noted, \textit{"With AI tools we can get solutions very quickly. Before ChatGPT, we had to search extensively for answers. Now we just ask, and it gives us the best solutions."} Practitioners also appreciated AI's continuous availability and rapid learning capacity, which offer practical benefits to their daily SE tasks (T1-ST2). As P3 states, \textit{"it's a twenty-four-seven available assistance".} However, practitioners acknowledge that AI has boundaries in adaptability and self-awareness (T1-ST1). P4 emphasizes this limitation:\textit{"it is effective, but we cannot trust it 100\%. Sometimes the tool gives wrong solutions."}

\noindent\textbf{T2.Socio-emotional gap as a collaboration barrier:}
Practitioners acknowledge that the socio-emotional gap creates significant obstacles in collaboration. AI models struggle to grasp user intent, and unstated requirements, leading to misalignment and inefficiency (T2-ST2). P1 underscores the importance of AI understanding the user intent: \textit{"Understanding is very important...I feel sometimes it doesn't understand what the user is trying to communicate with it. "} Furthermore, AI lacks the capacity to detect, understand, or respond to human emotional states in the manner that human collaborators can (T2-ST1). P6 observes: \textit{"it’s not 100\% affected by emotions like a human would be. It just reacts at a surface level..."} When AI does display emotional responses, practitioners perceive these as superficial mimicry without clear or genuine comprehension (T2-ST5). P5 pointed this out, \textit{"I think it’s already pretty good at this on a surface level, it can sound empathetic, even if it’s not genuine."} Furthermore, AI fails to maintain context across interactions, adapt to changes, or engage in sustained teamwork comparable to HH collaboration (T2-ST4). P1 highlights this discontinuity, \textit{"...it doesn't adapt to the changes...So then, instead of going to the next step, it goes to a different solution from the first...So then I feel there is never teamwork."} Given these socio-emotional limitations, practitioners express concern about AI surpassing humans while lacking the socio-emotional dimension (T2-ST3). P3 articulates this anxiety, \textit{"I'm a bit fearful because everything can be done with AI, the only thing that is missing is this emotional, social quotient."}

\noindent\textbf{T3.Human-specific collaboration traits:} Practitioners identify specific human qualities, including SEI, relationship skills, and collaborative drive, that they consider essential for meaningful work collaboration and superior to current AI capabilities. They emphasized that humans bring passion, drive, and shared commitment to collaborative work (T3-ST3). P1 observes that when people collaborate, \textit{"they are passionate about what they're doing, they both are trying to get something out there."} Practitioners also value the human ability to explain, articulate, and convey meaning effectively (T3-ST1). P5 emphasizes this unique capability: \textit{"knowing how to explain something to you in the way you’ll best understand it."} Humans also demonstrate genuine emotional comprehension and supportive behavior (T3-ST2) as highlighted by P3,\textit{"I expect my teammates should be more understanding in a situation, if one can't handle a situation, the person at the other side of the table should understand that and be able to help him."} Finally, practitioners value qualities such as building connections, establishing trust, and fostering cohesive team dynamics (T3-ST4). This preference for team-based qualities emerges clearly in P2's statement:\textit{"if we had a problem which is very critical and...about five or six people will be there to support us. So if I cannot find the reason, they will help... So it's teamwork."}

\noindent\textbf{T4.Cognitive value despite socio-emotional gap:} Despite recognizing socio-emotional limitations in AI tools, practitioners acknowledge the cognitive value AI tools provide, though performance remains inconsistent and context-dependent. Practitioners acknowledge that AI tools excel at providing speed, diverse perspectives, and iterative improvement in technical tasks, advantages that persist despite collaboration difficulties (T4-ST2). P3 illustrates this benefit: \textit{"For example, when I discuss something with a human teammate, my idea of the “perfect” solution today might change tomorrow because I may come up with a new perspective. But with AI, I can explore multiple options right away. I can ask whether there are alternative solutions and compare them instantly, which helps improve productivity."}However, practitioners struggle to evaluate AI's overall value, recognizing both benefits and limitations, with effectiveness being highly context-dependent and difficult to quantify. AI effectiveness varies based on task complexity, context, and user input quality, as acknowledged by the practitioners (T4-ST4). P6 estimates, \textit{"I would say about 70–80\% effective. It’s not 100\%."} AI performance heavily depends on prompt quality, problem clarity, and iterative refinement (T4-ST1). P6 pointed out, \textit{ "It doesn’t automatically generalize to other cases unless we specifically instruct it. Another problem is that if we provide one example, it only works for that case."} On a positive note, practitioners report that human-AI friction has decreased over time through learning and familiarity (T4-ST3). P5 mentioned this, \textit{"I find them pretty consistently helpful, and it's less often these days that it feels frustrating and much more often that it feels it's so good."}

\noindent\textbf{T5.Human agency and oversight:} Practitioners emphasize the importance and necessity of maintaining human control, oversight, and final decision-making authority in HAIC, highlighting the recognition of the value of human judgment. Practitioners emphasize the need for humans to retain the ultimate authority over important decisions in HAIC (T5-ST1). P4 explains this, \textit{"It doesn’t just give one solution, it provides options, and we choose the one that fits best."} Practitioners also show skepticism about AI reliability, which necessitates human control and verification in tasks and decision-making (T5-ST2). P3 states this, \textit{" I don't trust it. AI is always an outsider for me, so I don't expect anything from it." }

\begin{tcolorbox}[boxstyle,boxsep=2pt,left=3pt,right=3pt,top=3pt,bottom=3pt,before skip=6pt,after skip=6pt]
\textbf{Answer to RQ1:} Practitioners perceive a significant socio-emotional gap in HAIC and view AI as a cognitively valuable technical tool lacking socio-emotional awareness, contextual understanding, and collaborative continuity that cause collaboration barriers requiring human oversight.
\end{tcolorbox}
    
\subsection{RQ2: Evolution of HAIC Capabilities}
\label{subsec:RQ2}
Thematic analysis showed practitioners envision AI evolution towards collaborative partnership (T6) alongside technical and contextual enhancements (T7).

\noindent\textbf{T6.Desired AI enhancements for collaboration:} Practitioners envision AI developing more integrated, human-like collaborative capabilities, including role flexibility and authentic collaborative dimensions that can extend its current function as a passive tool. Practitioners emphasize their desire for AI to function as a teammate with diverse roles rather than as a passive tool(T6-ST1). P6 articulates this vision:\textit{"I don’t expect it to do everything, but I need it to work with me in a supportive way, like a teammate who helps me finish my task."} However, while practitioners prefer emotionally capable AI, they express concerns about authenticity, preferring genuine functional improvements over superficial emotional mimicry (T6-ST2). P1 point out, \textit{"AI itself is an artificial intelligence, which is trying to replicate human intelligence... we have a few people who fake emotions. We have an extra one with a fake emotion."} Practitioners highlighted that AI should learn independently, adapt contextually, and avoid mimicking human reasoning (T6-ST3). P5 advocates this autonomous learning:\textit{"if you just let them learn themselves, the solutions they find are so much better than if we try to be too hands-on and we try to hard-code too much into it."} Practitioners also believe that AI should challenge users appropriately, when necessary, while remaining receptive to feedback and behavioral changes (T6-ST4). P5 explains,\textit{"it mustn’t just be so flexible that it always agrees with the user and becomes manipulable. If you point out a bug but you’re actually wrong, you want the AI to tell you that you’re wrong."}Finally, practitioners also envision AI models to understand team objectives, organizational structures, and decision-making contexts to provide contextually appropriate guidance (T6-ST5). P5 points out, \textit{"So it could understand the hierarchy to be able to make decisions or advice in the right way."}

\noindent\textbf{T7.Technical and contextual enhancement needs:}
Practitioners prioritize advancements in AI that focus on technical reliability and trustworthiness rather than socio-emotional capabilities, with a vision of seamless technical integration that complements and augments, rather than replaces, human abilities. AI requires a better understanding of user needs, problem context, and current knowledge levels of the user to engage in meaningful collaboration (T7-ST1). P5 explains that AI should, \textit{"try to understand what your current understanding is and try to help you."} Practitioners also stress that AI should enhance rather than replace human intelligence and critical thinking (T7-ST2). P5 states that practitioners are, \textit{"losing their critical thinking. That might be super useful to actually start to incorporate into future releases to sort of prevent this brain rot that some people are worrying about that people are becoming too dependent on."} Practitioners also call for improvements in the information quality given to AI models, comparison capabilities, and trustworthiness in the current AI models. The vision of "collaborating like humans" appears to be about functional accuracy rather than EI, based on practitioners' input (T7-ST3). P4 states this priority, \textit{"Improving trust would be very useful. Right now, we cannot completely rely on it."}

\begin{tcolorbox}[boxstyle,boxsep=2pt,left=3pt,right=3pt,top=3pt,bottom=3pt,before skip=6pt,after skip=6pt ]
\textbf{Answer to RQ2:} Practitioners envision AI evolving into a more integrated collaborative partnership through adaptive learning while prioritizing technical reliability, contextual awareness, and human augmentation over synthetic socio-emotional intelligence, while humans maintain control and oversight.
\end{tcolorbox}

\subsection{Implications from results: Functional equivalents framework for effective HAIC}
\label{subsec:functionalequivalents}

\begin{figure*}[ht] 
    \centering
    \includegraphics[width=0.8\textwidth, height=0.25\textheight, keepaspectratio]{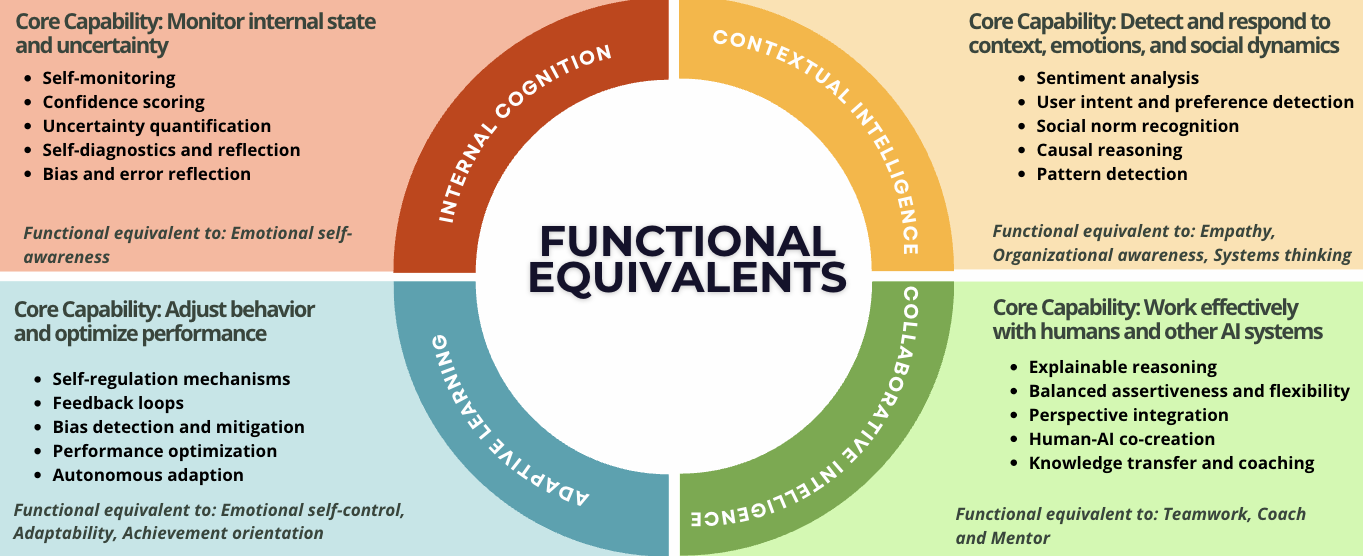}
    \Description{This diagram shows a framework listing the four key functional equivalents of SEI traits in humans that can enhance collaborative capabilities in HAIC through technical capabilities as preferred by SE practitioners.}
    \caption{Functional equivalents framework: Conceptualization of SEI traits for effective HAIC as AI technical capabilities that deliver the same collaborative outcomes as human SEI traits through distinct technical mechanisms}
    \label{functional_equivalent}
\end{figure*}

The findings highlight a key contradiction in HAIC within SE: while practitioners explicitly dismiss the need for artificial emotional and social capabilities and prioritize technical reliability and trustworthiness, they identify collaboration challenges from AI’s inability to perform roles typically supported by human SEI traits. This contradiction indicates that they do not necessarily expect emotionally intelligent AI, but rather functionally intelligent systems where AI can replicate the collaborative benefits of human SEI traits through distinct, technically driven mechanisms. This insight shows the importance of identifying \textbf{functional equivalents} (AI capabilities which serve the same collaborative purposes as human SEI traits without replicating emotional or social behavior) as shown in \autoref{functional_equivalent}. This framework responds to practitioners' concerns about artificial emotions while aligning with their stated preferences for enhanced collaboration, contextual understanding, and reliable technical performance given current AI capabilities. In short, the ESCI framework needs reinterpretation through a functional lens to map collaborative needs to technical capabilities that can enhance HAIC in SE contexts. To identify these functional equivalents, we analyzed each ESCI trait's collaborative outcome in human teams, then identified AI capabilities that could achieve similar outcomes through technical mechanisms. This mapping is based on practitioners' stated collaboration challenges and the capabilities they envision for effective HAIC. Thus, functional equivalents may be domain-specific, as what works for SE may not work for other domains (healthcare, education, creative work).  

\noindent\textbf{Internal cognition in AI models:} Practitioners' limited trust in AI tools is caused by the inability of AI tools to state their uncertainty and limitations. This lack of trust creates a fundamental collaboration barrier where the practitioners must independently verify all AI outputs~\cite{hamza2024human}, compromising efficiency gains. Furthermore, they raised concerns about AI competence without self-awareness, indicating the need for AI to recognize and communicate its limitations during specific tasks. Internal cognition in AI models addresses these challenges through technical self-awareness mechanisms such as confidence scoring, uncertainty quantification, and error detection~\cite{sanneman2022situation}. When AI models can monitor their internal states, determine the level of uncertainty in their responses through confidence scores, and perform self-diagnostics by being self-reflective about their own biases and errors ~\cite{jonker2024reflective}, they provide practitioners with the transparency needed to make informed trust decisions. This trait can be mapped to emotional self-awareness in human teammates (the ability in humans to recognize one's emotions and their effects). This capability achieves the same collaborative outcome as emotional self-awareness in humans (the ability to recognize one's emotions and their effects), but through statistical analysis rather than emotional self-awareness. This is not about AI "feeling" uncertain, but it is about AI reporting uncertainty through quantifiable metrics. When an AI system indicates "I am 60\% confident in this solution" or "This recommendation assumes X, which may not hold in your context," it functionally replicates what a self-aware human teammate does when saying "I am not entirely sure about this approach." Both enable the collaborator to balance their trust and determine appropriate decisions to be made. This functional equivalent directly addresses practitioners' need for improved trustworthiness without requiring the AI to possess human-like self-reflection.

\noindent\textbf{Contextual intelligence in AI models:} The main collaboration barrier practitioners identified is AI's failure in understanding the problem context, user intent, and the situational aspects that affect the collaboration for SE tasks. Contextual intelligence in AI models involves the ability of the model to detect trends, behaviors, and deeper insights while recognizing user preferences, intent, and situational factors to adjust its responses accordingly and remain intelligible. This includes sentiment analysis capabilities that detect emotional states from text or speech to adapt tone and responses appropriately~\cite{firdaus2021seprg,zhang2024impact}, understanding of social norms to provide organizationally appropriate guidance, and causal reasoning to understand interconnected factors in complex problems~\cite{kiciman2023causal}. These capabilities act as functional equivalents to multiple human SEI traits. Sentiment analysis and intent detection can achieve similar collaborative outcomes as empathy (sensing others' feelings and concerns), but through natural language processing. When AI detects frustration in a user's prompt or message and adjusts its explanations accordingly (not because it "feels" the frustration, but because it recognizes linguistic patterns), then it provides the same collaborative benefit a human teammate provides through empathetic response. Understanding organizational structures and group dynamics functionally replicates organizational awareness (understanding group emotions and power dynamics) through knowledge representation rather than social awareness. Similarly, causal reasoning and pattern detection in complex information give the collaborative benefits of systems thinking and pattern recognition, enabling AI to provide insights about interconnected factors without having human cognitive abilities. Practitioners are comfortable with this surface-level analysis as long as it functionally serves their collaboration needs, as they do not expect AI models to genuinely understand emotions but rather recognize contextual signals and respond appropriately.

\noindent\textbf{Adaptive learning in AI models:} Practitioners described AI's inability to maintain collaborative continuity and adapt to evolving contexts as a significant barrier in collaboration, which prevents the sustained teamwork practitioners value in human collaboration. Adaptive learning involves AI's ability to self-regulate outputs, adjust behavior to mitigate biases, and continuously refine performance through feedback mechanisms~\cite{zhao2025role}. This includes self-regulation to avoid harmful responses, which functionally maps to emotional self-control (keeping disruptive impulses in check) but operates through constraint optimization rather than impulse management. When an AI system adjusts its responses based on feedback to avoid biased or harmful outputs, it achieves the same collaborative benefit as a human teammate who regulates emotional reactions by maintaining professional interaction quality without requiring emotional experience. Practitioners also favored autonomous adaptation (adjusting to new situations and environments through self-learning~\cite{lu2024we}), which serves as a functional equivalent to human adaptability (flexibility in handling change). The key difference is mechanism: humans adapt through cognitive flexibility and experience, while AI adapts through parameter adjustment and pattern recognition. Both enable effective response to changing requirements. Furthermore, AI's capacity for continuous performance refinement addresses practitioners' recognition of context-dependent effectiveness through iterative improvement. This refinement capability functionally resembles achievement orientation (striving to improve and achieve excellence), though AI pursues optimization through algorithmic improvement rather than personal drive. Importantly, practitioners emphasized that adaptive learning should not create dependency that degrades human skills, so it must balance AI performance enhancement with human skill preservation.

\noindent\textbf{Collaborative Intelligence in AI models:} Despite viewing AI as a technical tool with clear boundaries, practitioners envision evolution toward a genuine collaborative partnership, contradicting current AI models that cannot engage in true work distribution or collaborative problem-solving. Practitioners value human collaboration specifically for passion, shared commitment, authentic communication skills, and team-based qualities such as trust and cohesion. The challenge is identifying how AI can contribute to collaborative outcomes without possessing these human qualities. Collaborative intelligence includes AI's ability to work effectively with humans and other AI systems through transparent reasoning, balanced perspectives, and support for human collaboration and co-creation~\cite{schleiger2024collaborative}. This includes providing clear explanations and justifications for decisions based on practitioners' needs. When AI makes its reasoning process visible and adapts explanations to user knowledge levels, it functionally replicates aspects of effective human communication without requiring a genuine understanding of the human experience. Practitioners desire AI that can challenge users while remaining receptive to feedback ~\cite{sharma2023towards}. This balanced assertiveness (knowing when to push back and when to adapt) functionally mirrors collaborative partnership in human teams. The functional equivalent of teamwork (collaborating toward shared goals) emerges when AI systems integrate seamlessly with human workflows, maintain context across interactions, and contribute substantively to joint problem-solving~\cite{schleiger2024collaborative}. Unlike human teamwork driven by social connection and shared identity, AI teamwork operates through technical integration and functional complementarity. Additionally, when AI tools support learning and skill development in collaboration, they exhibit capabilities similar to coaching and mentoring (supporting others' growth and abilities).

To summarize, AI need not imitate human SEI traits to be effective in collaboration for SE tasks, it only needs to achieve similar collaborative results through appropriate technical means.

\subsection{Validation of Functional Equivalents}
\label{subsec:validation}

Four software practitioners (VP1-VP4) were interviewed to assess the framework's perceived relevance for HAIC collaboration, not its demonstrated effectiveness. The participant details can be found in \autoref{tab:validationparticipants}. Validation findings mainly address current practical relevance rather than the validation of enduring principles.

\noindent\textbf{Internal cognition in AI models:}Interview data provided support for internal cognition while revealing key factors to consider when resorting to AI self-assessment. AI transparency through confidence scoring and uncertainty quantification can be valuable, especially for inexperienced practitioners (VP1). However, experienced practitioners prefer active verification over passive reliance in AI's uncertainty quantification, as they can independently identify errors(VP1). The reliability of the self-diagnostics and self-monitoring needs to be verified, as there are cases where AI can still produce error-prone codes even after error reflection and self-monitoring (VP2). However, this is similar to cases where human teammates are not always correct in their reasoning about self-awareness, but we value their self-awareness communication for better collaboration. The functional equivalence is not about perfect accuracy, but it is essential for enabling informed decision-making. Imperfect or unreliable confidence scoring and uncertainty quantification indicate where additional verification is needed. Despite this limitation, self-monitoring was viewed as beneficial since, in collaboration, humans ultimately oversee AI outputs (VP4). Source attribution was highlighted as a critical aspect in internal cognition (VP1) as practitioners could distinguish between AI-generated content and content from other sources. This would further support internal cognition for bias tracing, where we can assess bias in sources that AI uses rather than attributing bias in AI models themselves.

\noindent\textbf{Contextual intelligence in AI models:}Practitioners saw value in AI models having contextual intelligence to better adapt to user needs, understand project constraints, and provide more relevant and practical assistance in collaboration. When practitioners explicitly state contextual information such as deadline and time pressure, it could guide the AI responses (VP1, VP4). Even though current AI models are multimodal, they may struggle to detect implicit context without explicit communication (VP1) through sentiment analysis and preference detection. This highlights the difference in explicit and implicit context detection, especially in HAIC. However, it is to be noted that this resembles HH collaboration, where teammates require explicit communication of urgency, constraints, and preferences instead of automatically detecting all situational factors. Thus, the functional equivalent emphasizes both explicit and implicit contextual communication for optimal collaboration. Maintaining contextual boundaries across different domains of SE tasks could be challenging (VP2). This technical limitation of AI models needs to be addressed through better training and adaptive learning based on user feedback. Contextual intelligence could lead to more practical and achievable recommendations (VP4). For example, when coding within sprint deadlines or testing constraints, context-aware AI can suggest solutions that are actually feasible within the given timeframe rather than proposing scenarios impossible to achieve (VP4).

\noindent\textbf{Adaptive learning in AI models:} Practitioners confirm the relevance of learning and continuous refinement through feedback mechanisms. Adaptive learning must be flexible and task-dependent (VP1). For writing tasks, personalization is important, but for coding and technical tasks, correctness could be prioritized over personalization (VP1). The desired AI behavior also depends on different use cases and contexts (VP1). This is same as humans adapting to communication styles and contexts while maintaining technical standards. However, too much learning can lead to overfitting, where AI learns patterns so specifically that it becomes ineffective in generalized contexts (VP3). Additionally, excessive adaptation can create bias towards individual users and lead to learning incorrect patterns from the user (VP1). As SE tasks involve team-based activities, adaptive learning should evolve from individual to team level (VP2). For collaborative tasks such as documentation, AI needs to learn team-specific patterns, communication styles, and how the team reads documents to support collaborative continuity (VP2). Even with self-diagnostics and self-monitoring, AI models could lack awareness of when adaptation becomes counterproductive, requiring intervention from users (VP2) through feedback. This indicates that more advanced technical capabilities in AI models are needed to reach this level of self-regulation, in addition to feedback loops. This is similar to the human self-regulation, where team members often fail to recognize when their adaptability is inappropriate without explicit feedback or external intervention.

\noindent\textbf{Collaborative Intelligence in AI models:} Practitioners strongly valued the concept of explainable reasoning as an essential trait for collaborative intelligence. Explainable reasoning builds trust by being transparent on how AI arrives at a conclusion (VP1) and showing the analytical steps taken (VP4). It also enables iterative problem-solving by helping users understand what to prompt next for better responses (VP1), indicating that explainable reasoning helps knowledge transfer, a core collaborative function. This trait was compared to humans communicating project constraints to managers (VP3) to have better clarity in collaboration. However, there is an important distinction between transparency and truth (VP1). Even when AI explains reasoning clearly, it's essential to verify whether that explanation correctly reflects how the conclusion was reached (VP1). This is similar to human rationalization, where stated reasoning may differ from actual thought processes that could be intuitive or biased. This calls for human-AI co-creation that balances the biases and limitations of both parties. Practitioners pointed out that while explainable reasoning shows the "what," it doesn't necessarily teach the "how" (VP2), where users understand concepts, but they don't learn practical techniques. This indicates that AI tools enhance the capabilities of those who possess the technical expertise while failing to transfer that expertise to people who lack technical knowledge. This further widens the knowledge gap instead of closing it. Thus, knowledge transfer in collaborative intelligence is generally specific to the expertise level of the knowledge worker who is collaborating.

\section{Discussion}
This study examined software practitioners' perceptions of the socio-emotional gap in HAIC for SE (RQ1) and how HAIC capabilities could evolve to enable effective collaboration (RQ2). For RQ1, we explored how practitioners currently position AI in their workflows, what socio-emotional limitations they identify in HAIC, how these limitations compare to human collaborative capabilities, and the role of human oversight in managing this gap. For RQ2, we examined practitioners' visions for AI evolution and developed a functional equivalents framework. 

\noindent\textbf{Practitioners' perceptions of the socio-emotional gap in HAIC for SE:} Practitioners frame the socio-emotional gap as a functional gap in collaborative capabilities, not AI's failure to replicate emotions, recognizing AI cannot distribute work, negotiate responsibilities, or sustain problem-solving partnerships. While practitioners maintain clear ontological boundaries between human and AI capabilities~\cite{guzman2020ontological}, their technical knowledge shapes how they articulate required improvements: adaptive communication, contextual understanding, sustained collaboration, and transparent reasoning~\cite{sundar2020rise}. They understand communication and contextual issues arise from AI's inability to functionally interpret contextual situations that humans naturally process, not from AI failing to "feel" what users need. Similarly, the absence of emotional awareness reflects AI's inability to detect and respond appropriately to human emotional states rather than a need for AI to experience emotions. From their perspective, AI cannot be a true teammate because it lacks functional capabilities such as memory, adaptation, and contextual learning that characterize sustained human collaboration. This reveals that AI can be cognitively valuable while remaining collaboratively deficient.

Previous research positioned trust and anthropomorphization as essential for AI adoption in collaborative workflows~\cite{onnasch2021impact,epley2007seeing}. This study confirms the importance of trust but reveals that for technically proficient users, trust emerges from functional transparency and reliability rather than anthropomorphization. Domain expertise fundamentally shapes user responses to AI's socio-emotional attributes, as mental models of AI agency determine expectations~\cite{sundar2020rise}. Practitioners explicitly rejected artificial emotional displays, contrasting with research suggesting emotional expression enhances perceived humanness and social interactivity~\cite{zhang2024impact}. These findings diverge from other domains where emotional awareness enhances interaction quality~\cite{ferrada2024emotions} and positive emotional cues improve team dynamics~\cite{mallick2024you}. These divergences highlight domain-specific differences as SE contexts prioritize technical accuracy and reliability over emotional engagement, mainly due to practitioners' in-depth understanding of AI's technical capabilities and limitations. Building on these perceptions, the study explored how HAIC should evolve to address the functional gaps practitioners identified.

\noindent\textbf{Evolution of HAIC through functional equivalents:} The study indicates that the parallel to human emotional and social intelligence in AI models is about designing AI models with functionalities that can handle aspects such as user emotions, social context, and complex scenarios with sensitivity. This requires a multi-disciplinary design where the AI models are not only technically efficient but also built socially and ethically aligned for better collaborations. While technically proficient users view AI as intellectual teammates with less need for social traits, their emphasis on desirable SEI attributes indicates requirements for future collaboration as AI capabilities and landscape expand. AI needs to exhibit functionally equivalent emotional intelligence despite fundamentally different internal mechanisms, necessitating the need to include social awareness and emotional considerations in AI design, where the systems can engage in more adaptive and engaging collaborations. The findings provide a snapshot of current views on the perceptions and expectations of AI tools in terms of SEI, but it is important to know that the perspectives and expectations evolve as the AI technology evolves. Early AI models were regarded as tools rather than as collaborators. When the AI capabilities expanded into natural language processing and decision support, practitioners started expecting traits, namely reliability, transparency, and explainability, to mention a few. If AI models start exhibiting more human-like SEI traits in the future, such as social awareness and adaptive learning, then users will start expecting them to act in ways that mirror their human teammates. The study thus emphasizes that while current AI expectations are shaped by the current capabilities and exposure to the AI models, they are bound to change with the evolving landscape of AI technology. As AI capabilities evolve, the distinction between "functional equivalents" and "genuine" emotional intelligence may become harder to maintain or may become irrelevant.

The study aligns with research by Kolomaznik et al.~\cite{kolomaznik2024role}, which emphasized the need for AI systems to incorporate socio-emotional attributes for enhanced collaboration, but refines this by highlighting that technically proficient users require functional equivalents (technical capabilities achieving similar collaborative outcomes) rather than emotional replication. This difference is critical as it suggests that effective HAIC design must consider the technical expertise of users and domain-specific needs~\cite{cruz2024expert} rather than advocating universal socio-emotional enhancements in AI tools. Research on HAI co-learning~\cite{lu2024we} supports this study's finding that adaptive learning capabilities enable sustained collaboration. However, user acceptance of machine learning from interactions depends on perceived transparency and control~\cite{lee2023machine}, and the need for explainable adaptive mechanisms becomes crucial for maintaining effective HAIC. Similarly, research on explainable AI and situation awareness~\cite{sanneman2022situation} confirms practitioners' emphasis on transparent reasoning and contextual understanding as prerequisites for trust. This study extends this by explaining how these technical capabilities functionally replicate SEI traits such as self-awareness and empathy without requiring emotional capabilities that are authentic. 

Finally, research on reflective hybrid intelligence~\cite{jonker2024reflective} shows that self-reflective AI systems increase meaningful human control and empower human moral reasoning by providing comprehensible insights on moral blind spots, directly supporting the internal cognition category in the functional equivalents framework. Jonker et al. emphasized that self-reflective AI systems enable better human oversight, confirming practitioners' need for AI transparency about limitations. Research by Schleiger et al.~\cite{schleiger2024collaborative} supports the study's conceptualization of collaborative intelligence as distinct from human teamwork. They identified two-way task-related interaction between human and AI as key collaborative capabilities, aligning with practitioners' vision for AI that challenges appropriately while remaining receptive to feedback. The functional equivalents framework extends this by systematically mapping human collaborative traits to AI technical capabilities, as the essential aspect of collaboration is not equality between the actors but rather that they each contribute to shared output~\cite{schleiger2024collaborative}.

\section{Conclusion and Future Work}

This study explored the socio-emotional gap in HAIC within SE tasks, focusing on the expectations for SEI traits in human and AI teammates. The findings indicate that, while technically proficient users of current AI models perceived AI as a technical collaborator rather than a cognitive partner, they recognize the potential benefits of AI systems exhibiting certain functional equivalent traits of SEI traits, e.g., teamwork and adaptability. By mapping SEI traits to functional equivalents, we provide a structured direction for designing AI models that align with human expectations in collaborative settings. Our results support the idea that while current AI models may not possess human-like SEI traits, functional equivalent traits that address the collaboration challenges arising from this absence could improve HAIC efficiency. For SE practitioners, these findings highlight the need to prioritize HAI alignment by leveraging AI's functional capabilities to facilitate social and emotional engagement during collaboration. For AI developers, these findings highlight the need to develop AI models that align with the context and users' needs, systems that are adaptable, explainable, and cause minimal disruption in the current SE workflow. Future research should examine how collaboration dynamics and expectations differ when HH and HAI collaborate in different tasks, contexts, and domains. Investigating whether AI that adapts to users' expertise level could improve collaboration outcomes and whether HAIC expectations vary across different SE tasks could provide valuable insights for tailored collaboration interventions.

\bibliographystyle{ACM-Reference-Format}
\bibliography{sample-base}

\end{document}